\documentclass[proceedings]{stacs}
\stacsheading{2009}{99--110}{Freiburg}
\firstpageno{99}

\usepackage{bbm}
\usepackage{yhmath}
\usepackage{amssymb}

\usepackage[dvips]{epsfig}
\usepackage{graphics,psfrag}

\theoremstyle{plain}
\theoremstyle{definition}

\newcommand{\G}{\mathbb{G}}
\newcommand{\Z}{\mathbbm{Z}}
\newcommand{\N}{\mathbbm{N}}
\newcommand{\U}{\mathbbm{U}}
\renewcommand{\S}{\mathbbm{S}}
\newcommand{\supp}{\mathrm{supp}}
\newcommand{\A}{\mathcal{A}}
\newcommand{\B}{\mathcal{B}}
\newcommand{\E}{\mathcal{E}}

\newcommand{\T}{\mathbf{T}}
\newcommand{\Tfinal}{\mathbf{\Sigma}}
\newcommand{\Sub}{\mathbf{S}}
\newcommand{\W}{\mathbf{W}}

\newcommand{\s}{\sigma}

\newcommand{\TM}{\mathcal{M}}
\newcommand{\SFT}{\mathcal{SFT}}
\newcommand{\sofic}{\mathcal{S}ofic}
\newcommand{\shift}{\mathcal{S}}
\newcommand{\reshift}{\mathcal{RE}}
\newcommand{\fshift}{\mathcal{FS}}
\newcommand{\Uclass}{\mathcal{U}}

\newcommand{\Lang}{\mathcal{L}}

\begin{document}

% \title[short title]{title}
\title[An order on sets of tilings corresponding to an order on languages]{An order on sets of tilings corresponding to \\ an order on languages}

% \author[ref]{Short author}{Author}
\author[lab1]{N. Aubrun}{Nathalie Aubrun}
% \address[ref]{Address of authors with ref as reference}
\address[lab1]{Institut Gaspard Monge, Universit\'e Paris-Est Marne-la-Vall\'ee,
	\newline 77454 Marne-la-Vallée Cedex 2, France.}  %required
\email{nathalie.aubrun@univ-mlv.fr}  %optional

\author[lab2]{M. Sablik}{Mathieu Sablik}
\address[lab2]{ Laboratoire d'Analyse, Topologie, Probabilit\'e, Universit\'e de Provence,
	\newline 39, rue F. Joliot Curie, 13453 Marseille Cedex 13, France.}	%optional
\email{sablik@cmi.univ-mrs.fr}  %optional
%\urladdr{http://iml.univ-mrs.fr/$\sim$sablik/}  %optional

%\thanks{Thanks: \url{stacs.cls} and this sample file are derived from
%the files \url{lmcs.cls} and  \url{lmcs-smp.cls}.} %optional

%% mandatory lists of keywords and classifications:
% \titlecomment{OPTIONAL comment concerning the title, \eg, if a variant
% or an extended abstract of the paper has appeared elsewehere}
%%%%%%%%%%%%%%%%%%%%%%%%%%%%%%%%%%%%%%%%%%%%%%%%%%%%%%%%%%%%%%%%%%%%%%%%%%%

%% the abstract has to PRECEDE the command \maketitle:
%% be sure not to issue the \maketitle command twice!

\begin{abstract}
Traditionally a tiling is defined with a finite number of finite forbidden patterns. We can generalize this notion considering any set of patterns. Generalized tilings defined in this way can be studied with a dynamical point of view, leading to the notion of subshift. In this article we establish a correspondence between an order on subshifts based on dynamical transformations on them and an order on languages of forbidden patterns based on computability properties.

\end{abstract}

\maketitle

%% start the paper here:

\vspace{-0,4cm}

\section*{Introduction}\label{S:one}

Given a finite set of tiles $\A$ and a finite set of forbidden patterns $P$, a $d$-dimensional tiling is an element of $\A^{\Z^d}$ where the local conditions imposed by $P$ are satisfied at every point of $\Z^d$. This basic model captures geometrical aspect of computation~\cite{berger1966udp,robinson1971uan,Myers1974}. To establish structural properties of tilings, it is interesting to study the set of tilings which satisfy the conditions imposed by $P$~\cite{ballierstacs2008}. 

It is easy to generalize the usual notion of tiling considering an infinite set of forbidden patterns. A set of generalized tilings can be studied with a dynamical point of view with the notion of subshift~\cite{lind1995isd,kitchens1998sd}. In this theory, a set of usual tilings corresponds to a subshift of finite type. 

In dimension $1$, the class of subshifts of finite type is well understood. In particular, the language of a subshift of finite type is given by a local automaton~\cite{beal1993cs}. Given this result, it is natural to characterize subshifts whith a language given by a finite automaton. It is the class of sofic subshifts, which can all be obtained as a factor of a subshift of finite type~\cite{lind1995isd}. Thus, each sofic subshift is obtained by a dynamical transformation of a subshift of finite type. 

Multidimensional subshifts of finite type are not well understood. For example, it is not easy to describe their languages. Moreover, in addition to factors, there exist other types of dynamical transformations on multidimensional subshift: the sub-action of a $d$-dimensional tiling consists in taking the restriction of a tiling to a subgroup of $\Z^d$. Hochman showed that every $d$-dimensional subshift whose set of forbidden patterns is recursively enumerable can be obtained by sub-action and factor of a $d+2$-subshift of finite type~\cite{hochman2007drp}. 

This result suggests that a subshift can simulate another one, where the notion of simulation is given by operations on subshifts inspired by the dynamical theory. This involves different orders depending on the operations which are considered. In this paper, we present five types of operations: product, factor, finite type, sub-action and superposition. It is possible to formulate classic results with this formalism. Our main result (Theorem~\ref{EquivalenceOrdre}) establishes a correspondence between an order on subshifts based on dynamical transformations on them and an order on languages of forbidden patterns based on computability properties. 

The paper is organized as follows: Section~\ref{definition} is devoted to introduce the concepts of tiling and subshift. In Section~\ref{operation}, we present several operations on subshifts which allow to define the notion of simulation of a subshift by another one. Then, in Section~\ref{TuringMachine}, we define an important tool to define runs of a Turing machine with a sofic subshift. This tool is used to prove our main result in the last Section.

\section{Definitions}~\label{definition}
\vspace{-0,6cm}
	\subsection{Generalized tilings}

Let $\A$ be a finite alphabet and $d$ be a positive integer. A \emph{configuration} $x$ is an element of $\A^{\mathbbm{Z}^d}$. Let $\mathbbm{S}$ be a finite subset of $\mathbbm{Z}^d$. Denote $x_{\mathbbm{S}}$ the \emph{restriction} of $x$ to $\mathbbm{S}$. A \emph{pattern} is an element $p\in\A^{\mathbbm{S}}$  and $\mathbbm{S}$ is the \emph{support} of $p$, which is denoted by $\supp(p)$. For all $n\in\mathbbm{N}$, we call $\mathbbm{S}_n^d=[-n;n]^{d}$ the \emph{elementary support} of size $n$. A pattern with support $\mathbbm{S}_n^d$ is an \emph{elementary pattern}. We denote by $\mathcal{E}^d_{\A}=\cup_{n\in\mathbbm{N}} \A^{[-n;n]^{d}}$ the set of $d$-dimensional elementary patterns. A \emph{$d$-dimensional language} $\Lang$ is a subset of $\mathcal{E}^d_{\A}$. A pattern $p$ of support $\mathbbm{S}\subset\mathbbm{Z}^d$ \emph{appears} in a configuration $x$ if there exists $i\in\mathbbm{Z}^d$ such that for all $j\in\mathbbm{S}$, $p_j=x_{i+j}$, we note $p\sqsubset x$.

\begin{definition}
A \emph{tile set} is a tuple $\tau=(\A,P)$ were $P$ is a subset of $\mathcal{E}^d_{\A}$ called the \emph{set of forbidden patterns}.

A \emph{generalized tiling} by $\tau$ is a configuration $x$ such that for all $p\in P$, $p$ does not appear in $x$. We denote by $\T_{\tau}$ the set of generalized tilings by $\tau$. If there is no ambiguity on the alphabet, we just denote it by $\T_{P}$.
\end{definition}

\begin{remark}
If $P$ is finite, it is equivalent to define a generalized tiling by allowed patterns or forbidden patterns, the latter being the usual definition of tiling.
\end{remark}

             \subsection{Dynamical point of view : subshifts}

One can define a topology on $\A^{\mathbbm{Z}^d}$ by endowing $\A$ with the discrete topology, and considering the product topology on $\A^{\mathbbm{Z}^d}$. For this topology,  $\A^{\mathbbm{Z}^d}$ is a compact metric space on which $\mathbbm{Z}^d$ acts by translation via $\s$ defined by: 
$$\begin{array}{ccccc}
\s_{\A}^i:&\A^{\mathbbm{Z}^d} & \longrightarrow &\A^{\mathbbm{Z}^d}&\\
&x&\longmapsto & \s_{\A}^i(x)& \textrm{ such that } \s_{\A}^i(x)_u=x_{i+u} \ \forall u\in\Z^d.
\end{array}$$
for all $i$ in $\mathbbm{Z}^d$. This action is called the shift.

\begin{definition}\label{language}
A $d$-dimensional subshift on the alphabet $\A$ is a closed and $\s$-invariant subset of $\A^{\mathbbm{Z}^d}$. We denote by $\shift$ (resp. $\shift_d$, $\shift_{\leq d}$) the set of all subshifts (resp. $d$-dimensional subshifts, $d'$-dimensional subshifts with $d'\leq d$). 

Let $\T\subseteq \A^{\Z^d}$ be a subshift. Denote $\Lang_n(\T)\subseteq\A^{[-n;n]^d}$ the set of elementary patterns of size $n$ which appear in some element of $\T$, and $\Lang(\T)=\cup_{n\in\N} \Lang_n(\T)$ the \emph{language} of $\T$ which is the set of elementary patterns which appear in some element of  $\T$.
\end{definition}

It is also usual to study a subshift as a dynamical system~\cite{lind1995isd,kitchens1998sd}, the next proposition shows the link between both notions.

\begin{proposition}~\label{tiling-subshift}
The set $\T\subset\A^{\Z^d}$ is a subshift if and only if $\T=\T_{\Lang(\T)^c}$ where $\Lang(\T)^c$ is the complement of $\Lang(\T)$ in $\mathcal{E}_{\A}^d$.
\end{proposition}
%\begin{remark}\label{rem}
%One control that if  $\Lang$ verifies that ''$\forall p\in\Lang\subseteq\E^d_{\A}$ then for all pattern $p'\in\E^d_{\A}$ such that $p\sqsubset p'$ one has $p'\in\Lang$'' then $\Lang(\T_{\Lang})^c=\Lang$.
%\end{remark}
\begin{definition}
Let $\A$ be a finite alphabet and $\T\subset\A^{\Z^d}$ be a subshift.

The subshift $\A^{\Z^d}$ is the \emph{full-shift} associated to $\A$. Denote $\fshift$ the set of all full-shifts.

If there exists a finite set $P\subseteq\E^d_{\A}$ such that $\T=\T_P$ then $\T$ is a \emph{subshift of finite type}. Denote $\SFT$ the set of all subshifts of finite type. Subshifts of finite type correspond to the usual notion of tiling.

If there exists a recursively enumerable set $P\subseteq\E^d_{\A}$ such that $\T=\T_P$ then $\T$ is a \emph{recursive enumerable subshift}. Denote $\reshift$ the set of all recursive enumerable subshifts.
\end{definition}
\vspace{-0,5cm}
\section{Operations on tilings}\label{operation}

	\subsection{Simulation of a tiling by another one}

An \emph{operation} $op$ on subshifts transforms a subshift or a pair of subshifts into another one; it is a function $op:\shift\to\shift$ or $op:\shift\times\shift\to\shift$. We remark that a subshift $\T$ (resp. a pair of subshifts $(\T',\T'')$) and the image by an operation $op(\T)$ (resp. $op(\T',\T'')$) do not necessary have the same alphabet or dimension. An operation can depend on a parameter.% which allow to define classes of operation.

Let $Op$ be a set of operations on subshifts. Let $\Uclass\subset\shift$ be a set of subshifts. We define the \emph{closure} of $\Uclass$ under a set of operations $Op$, denoted by $\mathcal{C}l_{Op}(\Uclass)$, as the smallest set stable by $Op$ which contains $\Uclass$. 

We say that a subshift $\T$ \emph{simulates} a subshift $\T'$ by $Op$ if $\T'\in\mathcal{C}l_{Op}(\T)$. Thus there exists a finite sequence of operations chosen among $Op$, that transforms $\T$ into $\T'$. We note it by $\T'\leq_{Op}\T$. We remark that $\mathcal{C}l_{Op}(\T)= \{ \T'\ :\  \T'\leq_{Op}\T\}.$

	\subsection{Local transformations}

We describe three operations that modify locally the subshift.

\noindent$\bullet$ \textbf{Product $P$:}

Let $\T\subseteq\A^{\Z^d}$ and $\T'\subseteq\mathcal{B}^{\Z^d}$ be two subshifts of the same dimension, define: $$\phi_P(\T,\T')=\T\times\T'\subseteq(\A\times\mathcal{B})^{\Z^d}.$$
One has $\mathcal{C}l_{P}(\fshift)=\fshift\textrm{ and } \mathcal{C}l_{P}(\SFT)=\SFT.$

\noindent$\bullet$ \textbf{Finite type FT:}

These operations consist in adding a finite number of forbidden patterns to the initial subshift. Formally, let $\A$ be an alphabet, $P\subseteq\E_{\A}^d$ be a finite subset and let $\T\subseteq\A^{\Z^d}$ be a subshift. By Proposition~\ref{tiling-subshift}, there exists $P'$ such that $\T=\T_{P'}$. Define:
$$ \phi_{FT}(P,\T)= \T_{P\cup P'}.$$

If $P$ and $\T$ have not the same alphabet or the same dimension, put $\phi_{FT}(P,\T)= \T$. We remark that $\phi_{FT}(P,\T)$ could be empty if $P$ prohibits too many patterns. By $FT$, one lists all operations on subshifts which are obtained by $\phi_{FT}$.

By definition of subshift of finite type, one has $\mathcal{C}l_{FT}(\fshift)=\SFT$.

\noindent$\bullet$ \textbf{Factor F:}

These operations allow to change the alphabet of a subshift by local modifications. Let $\A$ and $\B$ be two finite alphabets. A \emph{morphism} $\pi:\A^{\Z^d}\to\B^{\Z^d}$ is a continuous function which commutes with the shift action (i.e. $\s^i\circ\pi=\pi\circ\s^i$ for all $i\in\Z^d$). In fact, such a function can be defined locally~\cite{hedlund1969eaa}: that is to say, there exists $\U\subset\Z^d$ finite, called \emph{neighborhood}, and $\overline{\pi}:\A^{\U}\to\B$, called \emph{local function}, such that $\pi(x)_i=\overline{\pi}(x_{i+\U})$ for all $i\in\Z^d$. Let $\T$ be a subshift, define:
$$\phi_F(\pi,\T)=\pi(\T).$$

If the domain of $\pi$ and $\T$ do not have the same alphabet or the same dimension, put $ \phi_{F}(\pi,\T)= \T$. By $F$, one lists all operations on subshifts which are obtained by $\phi_{F}$.

One verifies that $\mathcal{C}l_{F}(\SFT)\ne\SFT$.

\begin{definition}
A sofic subshift is a factor of a subshift of finite type. Thus, the set of sofic subshifts is $\sofic=\mathcal{C}l_{F}(\SFT)$.
\end{definition}

	\subsection{Transformation on the group of the action}

We describe two operations that modify the group on which the subshift is defined, thus we change the dimension of the subshift.

\noindent $\bullet$ \textbf{Sub-action SA:}

These operations allow to take the restriction of a subshift of $\A^{\Z^d}$ according to a subgroup of $\Z^d$. Let $\G$ be a sub-group of $\Z^d$ generated by $u_1,u_2,\dots,u_{d'}$ ($d'\leq d)$. Let $\T\subseteq\A^{\Z^d}$ be a subshift, define:
$$\phi_{SA}(\G,\T)=\left\{y\in\A^{\Z^{d'}}\ :\  \exists x\in\T \textrm{ such that } \forall i_1,\dots,i_{d'}\in\Z^{d'}, y_{i_1,\dots,i_{d'}}=x_{i_1u_1+\dots+i_{d'}u_{d'}}\right\}.$$

It is easy to prove that $\phi_{SA}(\G,\T)$ is a subshift of $\A^{\Z^{d'}}$. If $\T\subseteq\A^{\Z^d}$ and $\G$ is not a subgroup of $\Z^d$, put $ \phi_{SA}(\G,\T)= \T$. By $SA$, one lists all operations on subshifts which are obtained by $\phi_{SA}$.

One verifies that $\mathcal{C}l_{SA}(\SFT)\ne\SFT$ and $\mathcal{C}l_{SA}(\SFT)\ne\sofic$.

\begin{theorem}\label{stabRE}
$\mathcal{C}l_{SA}(\reshift)=\reshift$.
\end{theorem}

\noindent $\bullet$ \textbf{Superposition SP:}

These operations increase the dimension of a subshift by a superposition of the initial subshift. Let $d,d'\in\N^{\ast}$. Let $\G$ and $\G'$ be two subgroups of $\Z^{d+d'}$ such that $\G$ is isomorphic to $\Z^d$ and $\G\oplus \G'=\Z^{d+d'}$. Let $\T\subseteq\A^{Z^d}$ be a subshift, define:
$$\phi_{SP}(\G,\G',\T)=\left\{x\in\A^{\Z^{d+d'}}\ :\  \forall i\in\G',x_{i+\G}\in\T \right\}.$$

If $\T\subseteq\A^{\Z^d}$ and $\G$ is not isomorphic to $\Z^d$ or $\G\oplus\G'\ne\Z^{d+d'}$, put $ \phi_{SP}(\G,\G',\T)= \T$. By $SP$, one lists all operations on subshifts which are obtained by $\phi_{SP}$.

It is easy to verify that $\mathcal{C}l_{SP}(\SFT)=\SFT$.

With this formalism, the result of M. Hochman~\cite{hochman2007drp} can be written:
$$\mathcal{C}l_{F,SA}(\SFT)=\reshift.$$
More precisely, he proves that $\mathcal{C}l_{F,SA}(\SFT\cap \shift_{d+2})\cap\shift_{\leq d}=\reshift\cap\shift_{\leq d}.$

\section{Simulation of Turing machines by subshifts}\label{TuringMachine}

A Turing machine is a model of calculation defined by local rules. It seems natural to represent the runs of a machine by a 2-dimensional subshift: one dimension representing the tape and the other time evolution. But the main problem is that in general the Turing machine uses a finite part of the space-time diagram which is represented by the subshift. Robinson~\cite{robinson1971uan} proposes a self-similar structure to construct an aperiodic subshift of finite type of dimension $2$. In fact, it is also possible to use a general construction with substitutions due to Mozes~\cite{mozes1989tss}. This construction allows to give to the machine finite spaces on which it calculates independently. The problem is that we cannot control the entry of the Turing machine in view to recognize a configuration of a subshift. To obtain this property, Hochman~\cite{hochman2007drp} uses similar tools to construct a sofic subshift of dimension $3$ in order to to prove that $\mathcal{C}l_{F,S!
 A}(\SFT)=\reshift$. In this Section, we present a similar construction which is used to prove our main result in Section~\ref{MainResult}.

	\subsection{Substitution tilings}

Let $\A$ be a finite alphabet. A \emph{substitution} is a function $s:\A\rightarrow \A^{\U_k}$ where  $\U_k=[1;k]\times[1;k]$. We naturally extend $s$ to a function $s^n : \A^{\U_n} \rightarrow \A^{\U_{nk}}$ by identifying $\A^{\U_{nk}}$ with $(\A^{\U_{k}})^{\U_n}$.
Starting from a letter placed in $(1,1)\in\mathbbm{Z}^2$ and applying successively $s,s^k,\dots,s^{k^{n-1}}$ we obtain a sequence of patterns in $\A^{\U_{k^i}}$ for $i\in\{0,\dots, n\}$. Such patterns are called \emph{$s$-patterns}.

\begin{definition}
The subshift $\Sub_s$ defined by the substitution $s$ is
$$\Sub_s=\left\{ x\in\A^{\mathbbm{Z}^2}\ : \ \textrm{ every finite pattern of }x\textrm{ appears in a $s$-pattern} \right\}.$$
\end{definition}

	\subsection{A framework for Turing machines}

We now describe a family of substitutions $s_n$ defined on the alphabet $\{ \circ,\bullet\}$, which are used by M. Hochman~\cite{hochman2007drp} to prove $\mathcal{C}l_{F,SA}(\SFT)=\reshift$. For every integer $n$ the substitution $s_n$ is given by :
\begin{tiny}
$$
\begin{array}{ccccccc}
\circ
&
\longmapsto
& 
\begin{array}{cccc}
\circ & \dots & \circ & \circ\\
\vdots & \adots & \bullet & \circ\\
\circ & \adots & \adots & \vdots\\
\bullet & \circ & \dots & \circ 
\end{array}
&
\text{ \normalsize{ and } }
&
\bullet
&
\longmapsto
&
\begin{array}{cccc}
\circ & \dots & \circ & \bullet\\
\vdots & \adots & \bullet & \circ\\
\circ & \adots & \adots & \vdots\\
\bullet & \circ & \dots & \circ 
\end{array}
\end{array}
$$
\end{tiny}
where the patterns are of size $n\times n$. Let $\Sub_n$ be the tiling defined by substitution $s_n$.

These substitutions have good properties, in particular they are unique derivation substitutions and for this reason they verify~\cite{mozes1989tss}; one obtains:
\begin{proposition}
For every integer $n$, there exists a SFT  $\tilde{\Sub}_n$ and a letter-to-letter morphism $\pi_n$ such that $\Sub_n=\pi_n(\tilde{\Sub}_n).$
\end{proposition}

\begin{definition}
If $\T\subseteq  \A^{\Z^2}$ is a subshift, we define $\T^{(\uparrow)}$ by :
$$\T^{(\uparrow)}=\left\{x\in\A^{\Z^2} \ : \ \exists  y\in\T,\forall(i,j)\in\Z^2, x_{(i,j)}=y_{(i,j-i)} \right\}.$$

Notice that if $\T$ is an SFT, then $\T^{(\uparrow)}$ is also an SFT (just shift the forbidden patterns of $\T$ to get those of $\T^{(\uparrow)}$).
\end{definition}

We now work on the space $\mathbbm{Z}^3=\mathbbm{Z}e_1\oplus\mathbbm{Z}e_2\oplus\mathbbm{Z}e_3$ and we construct the SFT $\W_2$, $\W_3$ and $\W_5\subseteq \{ \circ,\bullet\}^{\mathbbm{Z}^3}$ defined by :
$$
\begin{array}{clcl}
x\in \W_2 \Longleftrightarrow
&
\left\{
\begin{array}{c}
\forall k\in\mathbbm{Z}, x_{|\mathbbm{Z}^2\times\{ k\}} \in \Sub_2^{(\uparrow)}\\
\forall u\in\mathbbm{Z}^3, x_u=x_{u+e_3} ~(*)
\end{array}
\right.
&
x\in \W_3 \Longleftrightarrow
&
\left\{
\begin{array}{c}
\forall j\in\mathbbm{Z}, x_{|\mathbbm{Z}\times\{ j\}\times\mathbbm{Z}} \in \Sub_3^{(\uparrow)}\\
\forall u\in\mathbbm{Z}^3, x_u=x_{u+e_2} ~(**)
\end{array}
\right.\\

x\in \W_5 \Longleftrightarrow
&
\left\{
\begin{array}{c}
\forall k\in\mathbbm{Z}, x_{|\Z^2\times\{ k\}} \in \Sub_5^{(\uparrow)}\\
\forall u\in\mathbbm{Z}^3, x_u=x_{u+e_3} ~(***)
\end{array}
\right.
&
&

\end{array}
$$

Let $x$ be a configuration of the subshift $\W_2\times \W_3\times \W_5 \subseteq (\{ \circ,\bullet \}^3)^{\Z^3}$. If we focus on the subshift $\W_3\times \W_5$, we can see rectangles whose corners are defined by the letter $(\bullet,\bullet)$ of $\{ \circ,\bullet \}^2$. These rectangles of size $5^n\times 3^m$ are spaces of calculation on which the Turing machine runs independently. Moreover the information brought by $\W_2$ gives the size of the entry pattern $p$ on each rectangle : scanning the base of a rectangle from left to right, the entry word is located between the left corner and the first symbol $\bullet$ due to $\W_2$ that occurs. This results are resumed in Proposition~\ref{frameworkTM}.

\begin{proposition}\label{frameworkTM}
The product $\W_2\times \W_3\times \W_5$ is a partition of the space into rectangles, in which each plane $\{ i\}\times\mathbbm{Z}^2$ is paved by rectangles of same width and height. Moreover if there is a $5^m\times 3^p$-rectangle in $(i,j,k)\in\Z^3$ with entry of size $2^n$, then there exists $i'$ and $i''$ such that there exists a $5^{m+1}\times 3^p$-rectangle in $(i',j,k)$ and a $5^m\times 3^{p+1}$-rectangle in $(i'',j,k)$ both with entry of size $2^n$. 
\end{proposition}

This result will be used in Section~\ref{recinclusion} to prove that, thanks to these arbitrary large rectangles, one can simulate a calculation with an arbitrary number of steps.

	\subsection{A $2$-dimensional sofic subshift}

We now explain how we can use the previously constructed framework to simulate a Turing machine by a subshift. First we recall the formal definition of a Turing machine.

\begin{definition}
Let $\mathcal{M}=(Q,\A,\Gamma,\sharp,q_0,\delta,Q_F)$ be a Turing machine, where :
\begin{itemize}
\item $Q$ is a finite set of states; $q_0\in Q$ is the initial state;
\item $\A$ are $\Gamma$ are two finite alphabets such that $\A\subsetneq\Gamma$;
\item $\sharp\notin\Gamma$ is the blank symbol;
\item $\delta : Q\times\Gamma\to Q\times\Gamma\times\{\leftarrow,\cdot\, ,\rightarrow\}$ is the transition function;
\item $F\subset Q_F$ is the set of final states.
\end{itemize}
\end{definition}

We can describe its behaviour with a set of 2-dimensional patterns. First dimension stands for the tape and second dimension for time evolution. For example the rule $\delta(q_1,x)=(q_2,y,\leftarrow)$ will be coded by :

$$
\begin{array}{|c|c|c|}
\hline
 (q_2,z) & y & z'\\
\hline
 z & (q_1,x) & z'\\
\hline
\end{array}
$$

Denote by $P_{\TM}$ the set of forbidden patterns constructed according to the rules of $\TM$. One can consider the subshift of finite type $\T_{P_{\TM}}$ where each local pattern corresponds to calculations of the machine $\TM$. Then thanks to a product operation we superimpose these calculations on the framework, with the following finite conditions :
\begin{itemize}
\item condition \textbf{Init} : to copy out the entry word ;
\item condition \textbf{Head} : the initial state $q_0$ appears on every rectangle bottom left corner and only here;
\item condition \textbf{Stop} : when a side of a rectangle is reached by the head of the machine, the calculation stops and if necessary the tape content is just copied out until the top of the rectangle;
\item condition \textbf{Final} : when a final state is reached, the tape content is just copied out for next steps of calculation until the top of the rectangle.
\end{itemize}

Define $\T_{\TM}$ the subshift:
$$\T_{\TM} = \phi_{FT}\hspace*{-0.25mm}\left(\hspace*{-1mm}\{\textbf{Init},\textbf{Head},\textbf{Stop},\textbf{Final} \},\A^{\Z^3}\hspace*{-2mm}\times\hspace*{-1mm}\left(\W_2\hspace*{-1mm}\times\hspace*{-1mm} \W_3\hspace*{-1mm}\times\hspace*{-1mm} \W_5\right) \hspace*{-1mm}\times\hspace*{-1mm} \phi_{SP}(\Z e_2\oplus\Z e_3,\Z e_1,\T_{P_{\TM}})\hspace*{-1mm}\right)\hspace*{-1mm}.$$

By stability of the class of subshifts of finite type by $SP$, $\T_{\TM}$ is a subshift of finite type up to a letter-to-letter morphism; thus $\T_{\TM}\in\sofic$. For all $i\in\Z$, in the plane $\{i\}\times\Z^2$, it is possible to find rectangles of size $5^m\times 3^p$ arbitrary large and an entry of size $2^n$ also arbitrarily large. On each rectangle, thanks to the conditions $P_{\TM}$, we can observe the evolution of the Turing machine $\TM$.

\begin{remark}
The construction described here only works for usual Turing machines. In Section~\ref{recinclusion} we explain how to add finite conditions on the subshift $\T_{\TM}$ if $\TM$ is a Turing machine with oracle. 
\end{remark}

\section{Study of the semi-order $\leq_{P,F,FT,SA,SP}$}\label{MainResult}

In this section we focus on the five operations described previously. Our aim is to study the semi-order $\leq_{P,F,FT,SA,SP}$.

	\subsection{A semi-order on languages}

A Turing machine with \emph{semi-oracle} is a usual machine with a special state $q_?$ and an oracle tape. The behaviour of a Turing machine with semi-oracle $\Lang$, where $\Lang$ is a language, is the following : the machine reads an entry pattern $p$ and writes a pattern on the oracle tape, until the state $q_?$ is reached. If the pattern written on the oracle tape is in $\Lang$ then the machine stops, else it keeps on calculating.

We define a semi-order on languages :
$$\Lang\preceq \Lang'\Longleftrightarrow \exists \TM^{\Lang'}\text{ a Turing machine with semi-oracle }\Lang'\text{ such that }dom(\TM^{\Lang'})=\Lang,$$
where $dom(\TM)$ is the domain of the machine $\TM$, that is to say the set of entry words on which $\TM$ stops. We refer to~\cite{rogersjr1987trf} for definitions and properties of similar semi-orders on languages based on computability.

\begin{prop}\label{semi-order}
$\preceq$ is a semi-order.
\end{prop}

Consider the equivalence relation $\Lang\approx \Lang'$ if and only if $\Lang\preceq \Lang'$ and $\Lang'\preceq \Lang$. This equivalence relation defines classes of languages, and we can compare them within the semi-order. For instance, the class of recursively enumerable languages is the smallest for this semi-order. We have $\emptyset\approx \Lang$ for every recursively enumerable language $\Lang$.

	\subsection{Closure theorem: }

The semi-order on languages defined by semi-oracle Turing machines corresponds to a semi-order on  subshifts:

\begin{theorem}\label{EquivalenceOrdre}
Let $\T$ be a subshift, one has:
$$\mathcal{C}l_{P,F,SA,SP,FT}(\T)=\left\{ \T_{\Lang} : \Lang\preceq \Lang(\T)^c\right\}.$$
Or equivalently, if $\T'$ and $\T''$ are two subshifts of dimension $d'$ and $d''$, one has:
$$\T' \leq_{P,F,FT,SA,SP}\T'' \Longleftrightarrow \Lang(\T')^c\preceq \Lang(\T'')^c.$$
\end{theorem}

%%%%%%%%%%%%%%%%%%%%%%%%%%%%

\subsubsection{Direct inclusion} Put $\Lang=\Lang(\T)^c$. To show $\mathcal{C}l_{P,F,SA,SP,FT}(\T)\subseteq\{ \T_{\Lang'} : \Lang'\preceq \Lang\}$, it is sufficient to show the stability of $\{ \T_{\Lang'}: \Lang'\preceq \Lang\}$ by all the operations. Let $\Lang_1\subseteq\E^{d_1}_{\A_1}$ and $\Lang_2\subseteq\E^{d_2}_{\A_2}$ be two languages such that $\Lang_i\preceq \Lang$ for $i\in\{1,2\}$. Thus, for $i\in\{1,2\}$, there exists a Turing machine $\TM_i$ with semi-oracle $\Lang$ whose domain is exactly $\Lang_i$.

$\bullet$ \textbf{Stability under product:}
Let $\T'=\phi_P(\T_1,\T_2)$, so $\T'=\T_{\Lang'}$ with $\Lang'=\Lang_1\times\E^{d_2}_{\A_2}\cup\E^{d_1}_{\A_1}\times\Lang_2$. The language $\Lang'$ could be the domain of a Turing machine $\TM'$ with semi-oracle $\Lang$. It suffices to simulate the two Turing machines $\TM_1$ and $\TM_2$ (each machine runs during one step successively) on each coordinate of a pattern of $\Lang'$. Thus $\Lang'\preceq\Lang$.

$\bullet$ \textbf{Stability under finite type:}
Let $\T'=\phi_{FT}(P,\T_{\Lang_1})$. Since $P$ is finite, one has $\Lang_1\cup P\preceq\Lang_1\preceq\Lang$ and $\T'=\T_{\Lang_1\cup P}$.

$\bullet$ \textbf{Stability under factor map:} Let $\T'=\phi(\pi,\T_{\Lang_1})$ where $\pi:\A_1^{\Z^{d_1}}\rightarrow \B^{\Z^{d_1}}$ is a morphism of neighborhood $\S_n^{d_1}$ and local function $\overline{\pi}$. One has $\T'=\T_{\Lang'}$ where $\Lang'=(\overline{\pi}(\Lang_1^c))^c$. Moreover, one has $\Lang'\preceq\Lang_1$. Indeed, if $p\in\E^{d_1}_{\B}$, we simulate the machine $\TM_1$ on all pattern $p'\in\A^{supp(p)+\mathbbm{\S}_n^{d_1}}$ such that $\overline{\pi}(p')=p$, running successively one step for each pattern. 

$\bullet$ \textbf{Stability under sub-action:} Let $\T'=\phi_{SA}(\G,\T_{\Lang_1})\subseteq\A_1^{\Z^{d'}}$ where $\G$ is a subgroup of $\Z^{d_1}$ of dimension $d'\leq d_1$. We consider the language $\Lang'\subseteq\E_{\A_1}^{d_1}$ which is the domain of the Turing machine $\TM'$: on a pattern $p\in\E_{\A_1}^{d'}$ of support $\U$, a Turing machine $\TM'$ simulates successively $\TM_1$ on every entry word of support $[-n;n]^{d_1}$ which completes $p$ in $\E^{d_1}_{\A_1}$ where $[-n;n]^{d_1}$ is the minimal support which contains $\U$ embedded in $\G$. Thus $\Lang'\preceq\Lang_1$, moreover $\T'=\T_{\Lang'}$. This is exactly the same principle as in the proof of Theorem~\ref{stabRE}.

$\bullet$ \textbf{Stability under superposition:} Let $\T'=\phi_{SP}(\G,\G',\T_{\Lang_1})$ where $\G$ is isomorph to $\Z^{d_1}$ and $\G\oplus \G'=\Z^{d_1+d}$. Let $\Lang'\subseteq\E^{d_1+d}_{\A_1}$ be the language where each pattern $p$ is the superposition of patterns $p_1,\dots, p_{d}\in\E_{\A_1}^{d_1}$ and there exists $i\in\{1,\dots,d\}$ such that $p_i\in\Lang_1$. Thus $\Lang'\preceq\Lang_1$ and $\T'=\T_{\Lang'}$.

%%%%%%%%%%%%%%%%%%%%%%%%%%%%%%%%%%%%%%%%%%%%%%%%%%%%%%%%%%%%%%%%%%%%%%%%%%%%%%%%%%%%%%%%%%%%%%%%%%%%%%%%%%%%%%%%%

\subsubsection{Reciprocal inclusion}\label{recinclusion}

Let $\T\subseteq \A^{\Z^d}$be a subshift; define $\Lang=\Lang(\T)^c\subseteq\E_{\A}^d$. Let  $\Lang'\subseteq\E_{\B}^{d'}$ be a language such that $\mathcal{L'}\preceq\Lang$.  We want to prove that $\T_{\Lang'}\in\mathcal{C}l_{P,F,SA,SP,FT}(\T)$.  

%Without loss of generality, we can suppose that $\Lang$ is a set of one-dimensional patterns. If it is not, we apply the following transformation : since there exists a recursive bijection from $\mathbbm{Z}^d$ to $\mathbbm{Z}$, every pattern in $\mathcal{E}_\A^d$ can be seen as a pattern in $\mathcal{E}_\A^1$. So there is a bijection between sets of forbidden patterns in $\mathcal{E}_\A^d$ and sets of forbidden patterns in $\mathcal{E}_\A^1$. Thus we can assume that $\Lang$ is a one-dimensional language.

Here, we assume that $\Lang$ and $\Lang'$ are one-dimensional languages, but the proof can be adapted to the general case. We explain how to construct the subshift $\T_{\Lang'}$ thanks to operations $P,F,FT,SA$ and $SP$ applied on $\T=\T_\Lang$.

Since $\Lang'\preceq \Lang$ there exists a Turing machine $\TM$ with semi-oracle $\Lang$ such that $dom(\TM)=\Lang'$. We transform this Turing machine so that it only takes in input patterns of support $[0,2^{n-1}]$ (because checked patterns are given by $\W_2$) and at the moment when the state $q_?$ is reached, the word written on the oracle tape is copied out in the alphabet $\widetilde{\A}$, which is simply a copy of $\A$, then again copied out in the alphabet $\A$ once the oracle has given its answer.

We first list auxiliary subshifts that we need to construct $\T_{\Lang'}$ :

\begin{itemize}
\item the original subshift $\T_\Lang$ written in the copy of $\A$: $\widetilde{\T}_\Lang\subseteq \widetilde{\A}^\Z$ will simulate the oracle;
\item Turing machine $\TM$ is coded by a subshift of finite type $\T_{\TM}\subseteq \mathcal{O}^{\Z^2}$, where $\mathcal{O}$ is an alphabet that contains at least $\A$, $\widetilde{\A}$ and $\B$;
\item the framework for this Turing machine will be given by $\W_2$, $\W_3$ and $\W_5$ defined in Section~\ref{TuringMachine}; they are defined on the alphabet $\{ \bullet,\circ \}$ and are subshifts of finite type up to a letter-to-letter morphism.
\end{itemize}

%\vspace{-0,2cm}
\paragraph{\textbf{Construction of $\T_{\Lang'}$}}
The principle is to construct $\Tfinal\in\mathcal{C}l_{P,F,SA,SP,FT}(\T_\Lang)$ a $4$-dimensional subshift on the alphabet $\mathcal{C}=\A\times\widetilde{\A}\times\B\times\{\bullet;\circ\}^3\times\mathcal{O}$. Denote $(e_1,e_2,e_3,e_4)$ the canonical basis of $\mathbbm{Z}^4$. We need these four dimensions for different reasons :
\begin{itemize}
\item the subshift $\T_{\mathcal{L'}}$ will appear on $\Z e_1$;
\item thanks to $\Z e_1\oplus\Z e_2\oplus\Z e_3$, we construct a framework for $\TM$, so that every rectangle of this framework is in a plane $\{ i\}\times\Z\times\Z\times\{ k\}$ where $i,k\in\Z$;
\item on $\Z e_4$ we have the oracle simulated by $\widetilde{\T}_\Lang$.
\end{itemize}
%\vspace{-0,3cm}
\paragraph{Step 1 :} First notice that changing $\T_\Lang$ into $\widetilde{\T}_\Lang$ only requires a letter-to-letter morphism. Then we construct $\tilde{\W}=\phi_{SP}(\Z e_4, \Z e_1\oplus\Z e_2\oplus\Z e_3,\widetilde{\T}_\Lang)$ to place $\widetilde{\T}_\Lang$ in a $4$-dimensional subshift. We finally add through a product operation $P$ all letters from $\mathcal{C}$ : $\W=\tilde{\W}\times(\A\times\B\times\{\bullet;\circ\}^3\times\mathcal{O})^{\Z^4}$ so that $\W \in\mathcal{C}l_{P,F,SP}(\T_\Lang)\cap\mathcal{C}^{\Z^4}$.

\paragraph{Step 2 :} We want $\T_{\Lang'}$ to appear on $\mathbbm{Z}e_1$. Simulations of the Turing machine $\TM$ will take in input a word written on $\mathbbm{Z}e_2$. So we need to copy out $\mathbbm{Z}e_1$ on $\mathbbm{Z}e_2$ so that these simulations apply to what will be the subshift $X_{\Lang'}$. We get to it with the finite condition :
$$\forall x\in \mathcal{C}^{\mathbbm{Z}^4}, \forall u\in\mathbbm{Z}^4, x_u=x_{u+e_1-e_2}.$$
We also want to keep accessible all along the simulation the entry word of every rectangle of the framework. To do that we add the finite condition :
$$\forall x\in \mathcal{C}^{\mathbbm{Z}^4}, \forall u\in\mathbbm{Z}^4, x_u=x_{u+e_3}.$$
We thus obtain a subshift $\W'\in\mathcal{C}l_{P,F,SP,FT}(\T_\Lang)$.
%\vspace{-0,2cm}
\paragraph{Step 3 :} Then we add to $\W'$ a framework for the Turing machine. We construct $W_{\textrm{rect}}\subseteq\{\bullet,\circ \}^{\Z^3}$ an auxiliary subshift of finite type up to a letter-to-letter morphism, containing well-chosen rectangles. Denote $F_i$ the finite type condition that ensures $\forall x\in\{\bullet,\circ \}^{\Z^3} , \forall u\in\mathbbm{Z}^3, x_u=x_{u+e_i}$. As in Section~\ref{TuringMachine}, we define:
\begin{itemize}
\item  $\W_2=\phi_{FT}(F_3, \phi_{SP}(\Z e_1\oplus\Z e_2,\Z e_3\oplus\Z e_4, \Sub_2^{(\uparrow)}))$;
\item  $\W_5=\phi_{FT}(F_3, \phi_{SP}(\Z e_1\oplus\Z e_2,\Z e_3\oplus\Z e_4, \Sub_5^{(\uparrow)}))$;
\item  $\W_3=\phi_{FT}(F_2,\phi_{SP}(\Z e_1\oplus\Z e_3,\Z e_2\oplus\Z e_4,\Sub_3^{(\uparrow)}) )$.
\end{itemize}

The rectangles are obtained in $\tilde{\W}_{\textrm{rect}}=\W_2\times \W_5\times \W_3$. Each rectangle of length $5^m$ given by $\W_5$ knows the length of its input $2^n$ given by $\W_2$. Thus we can simulate the Turing machine on words of length $2^n$, on a tape of length $5^m$ and simulations are bounded by $3^p$ steps of calculation. Up to a letter-to-letter morphism, $\tilde{\W}_{\textrm{rect}}$ is a subshift of finite type, so there exists a finite set of patterns $F_{\textrm{rect}}$ and a morphism $\pi_{\textrm{rect}}$ such that $\tilde{\W}_{\textrm{rect}}=\pi_{\textrm{rect}}(\T_{F_{\textrm{rect}}})$. We add this framework to $\W'$ via $\W_{\textrm{rect}}=\pi_{\textrm{rect}}(\phi_{FT}(F_{\textrm{rect}},\W'))$ so that we have $\W_{\textrm{rect}} \in\mathcal{C}l_{P,F,SP,FT}(\T_\Lang)$.

\paragraph{Step 4 :} We add the behaviour of $\TM$ in rectangles of $\W_{\textrm{rect}}$ but for the moment we do not take into consideration calls for oracle. As in Section~\ref{TuringMachine}, we consider the finite conditions $P_{\TM}$ given by the rule of $\TM$ and the conditions $P_{\textrm{calc}}=\{\textbf{Init}, \textbf{Head},\textbf{Stop},\textbf{Final}\}$ which control the interaction of the head of $\TM$ with the rectangles. For the moment every time the machine calls the oracle it keeps on calculating. Thus $\W_{\TM}=\phi_{FT}(P_{\TM}\cup P_{\textrm{calc}},\W_{\textrm{rect}}) \in \mathcal{C}l_{P,F,FT,SP}(\T_\Lang)$.

\paragraph{Step 5 :} To simulate the oracle, we add finite type conditions to ensure that during a calculation, when the machine calls for the oracle in $(i,j,k,l)\in\Z^4$, the pattern $p\in\tilde{\A}^n$ on which the oracle is called coincides with the pattern in $\Z e_4$ between $(i,j,k,l)$ and $(i,j,k,l+n)$. These new allowed patterns look like :
$$
\begin{array}{ccccc}
 \uparrow_{e_4} & 
\begin{array}{|c|c|}
\hline 
\tilde{a} & .\\
\hline
\tilde{b} & \tilde{a}\\
\hline
\end{array}~, 
 &
\begin{array}{|c|c|}
\hline
\tilde{a} & .\\
\hline
(q_?,\tilde{b}) & \tilde{a}\\ 
\hline
\end{array}
\\
 & \rightarrow_{e_2} & 
\end{array}
$$
However, these conditions are only valid in the interior of a rectangle. We denote these finite type conditions by $F_{\textrm{oracle}}$. Then we have $\W_{\TM_{\textrm{oracle}}}=\phi_{FT}(F_{\textrm{oracle}},W_{\TM}) \in \mathcal{C}l_{P,F,SP,FT}(\T_\Lang)$.

\paragraph{Step 6 :} In order to avoid dependence problems between different calculations, each configuration of $\T_\Lang$ that appears on $\mathbbm{Z}^4$ is used for the same calculation, thanks to the finite type condition :
$$\forall x\in \mathcal{C}^{\mathbbm{Z}^4}, \forall u\in\mathbbm{Z}^4, x_u=x_{u+e_1+e_4}.$$
Finally we consider the final state $q_{\textrm{stop}}$ as a forbidden pattern and we denote by $\Tfinal$ this subshift. We have $\Tfinal\in \mathcal{C}l_{P,F,SP,FT}(\T_\Lang)$.

We simulate the running of the Turing machine $\TM$ on a pattern $p\in\E^1_{\B}$ of length $2^n$. As soon as $\TM$ calls for the oracle, we compare the word on which the oracle is called and the word on $\Z e_4$. If the two words coincide then $\TM$ keeps on calculating, else it comes to the final state $q_{\textrm{stop}}$. If the machine cannot terminate its calculation within the time given by the rectangle, Proposition~\ref{frameworkTM} ensures that we can find a larger rectangle in which the machine will calculate on the same entry word.

\vspace{0,4cm}
The following picture resumes the behaviour of the machine $\TM$ on the framework :

\begin{center}
\psfrag{e1}{$e_1$}
\psfrag{e2}{$e_2$}
\psfrag{e3}{$e_3$}
\psfrag{e4}{$e_4$}
\psfrag{m}{$p$}
\psfrag{q?}{$q_?$}
\psfrag{qstop}{$q_{\textrm{stop}}$}
\includegraphics[width=7cm]{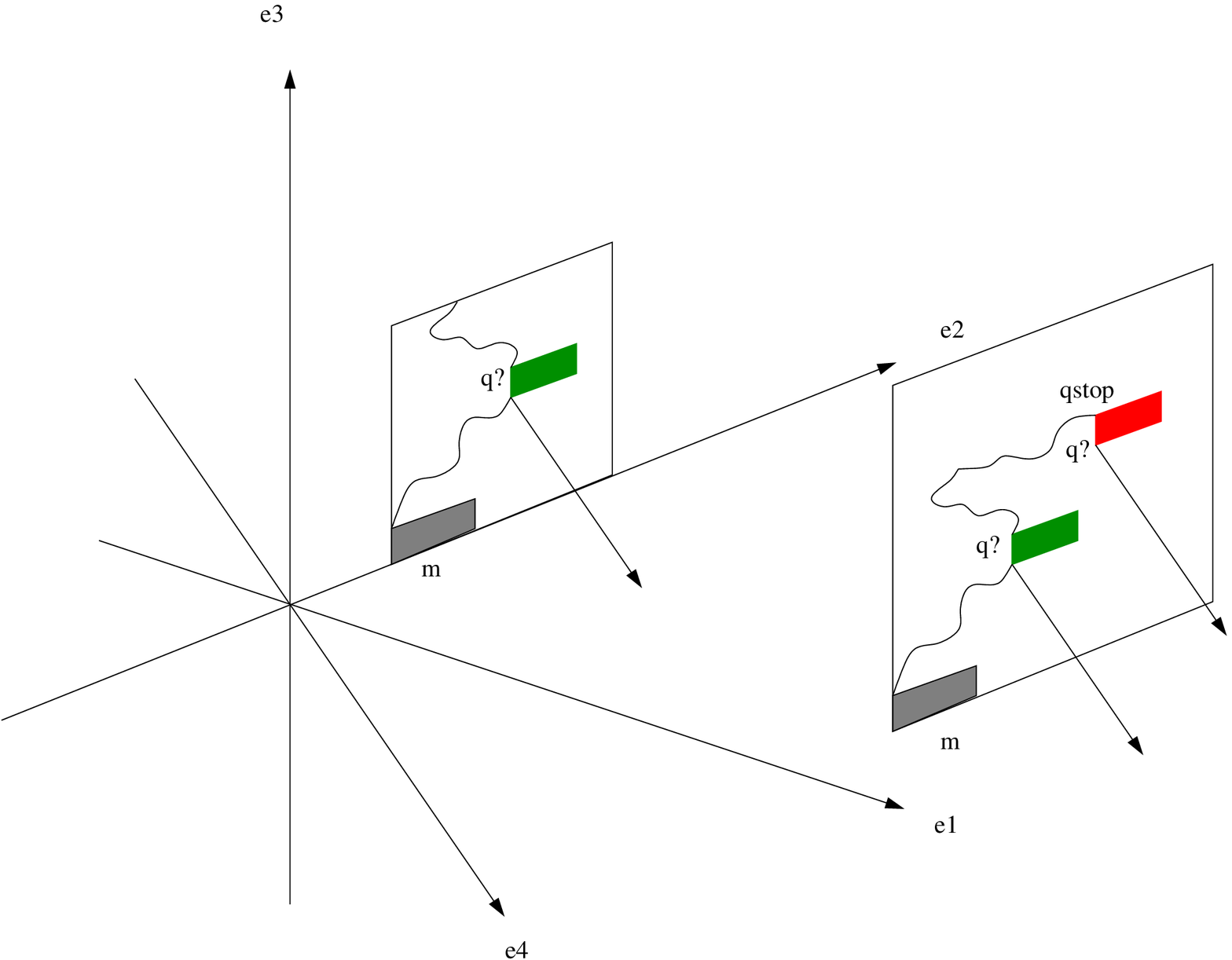}
\end{center}

\paragraph{\textbf{Proof that this construction works}}
We now prove that $\phi_{SA}(\Z e_1,\Tfinal)$, the projection of $\Tfinal$ on $\mathbbm{Z}e_1$ is $\T_{\Lang'}$, up to a morphism that just consists in keeping information about $\mathcal{B}$.

\paragraph{Proof of $\phi_{SA}(\Z e_1,\Tfinal) \subseteq \T_{\Lang'}$:}

Let $y\in\Tfinal$, we prove that $x=y_{|\Z e_1}\in \T_{\Lang'}$. It is sufficient to prove that every pattern in $x$ is not in $\Lang'$. Let $p$ be a pattern in $x$; it is a sub-pattern of a certain $p'\sqsubset x$ where $p'$ is chosen such that it is of length $2^n$. By construction of $\W_{\textrm{rect}}$ there exists $t,s\in\N$ arbitrary large such that there exists a rectangle of size $5^s\times3^t$ with the entry word $p'$. Since $y\in\Tfinal$, in every rectangle the calculation of the machine $\TM$ on the word $p'$ does not reach the final state $q_{\textrm{stop}}$. Since these rectangles are arbitrarily large, we can conclude that the machine $\TM$ never reaches $q_{\textrm{stop}}$. It means that $p'\notin \Lang'$, thus $p\notin \Lang'$.

\paragraph{Proof of $\T_{\Lang'} \subseteq \phi_{SA}(\Z e_1, \Tfinal)$:}

Let $x\in \T_{\Lang'}$, we construct $y\in\mathcal{C}^{\Z^4}$ such that  $y\in \Tfinal$ and $y_{|\Z e_1}=x$. To insure that $y\in\Tfinal$ we just need to check that for all $(i,j,k)\in\Z^3$, we can impose that $y_{| \{ i \} \times\{ j \}\times \{ k \} \times \Z} \in \T_\Lang$ while the calculations of $\TM$ in the rectangles containing any $(i,j,k,l)$ do not reach the state $q_{\text{\textrm{stop}}}$.

Let us now focus on a specific rectangle of the framework, on which the machine $\TM$ calculates on a pattern $p$ of size $2^n$ that appears in $x$. Since $p$ appears in $x$, $p\notin \Lang'$ so the machine $\TM$ loops on the entry $p$. It means that every time the calculation of $\TM$ on $p$ calls for the oracle on a pattern $p'$, $p'$ is not in $\Lang$. Since $\Lang=\Lang(\T)^c$, for all pattern $p'$ on which the oracle is called, there exists a configuration $z\in \T_\Lang$ such that $z_{|[0;|m'|-1]}=p'$. Thus we complete $y$ on the following way :
\begin{itemize}
\item[-] if in $(i,j,k)\in\Z^3$ the calculation of $\TM$ calls for the oracle on a pattern $p'$, then $y_{| \{ i \} \times\{ j \}\times \{ k \} \times \Z}=z$ previously constructed;
\item[-] if the oracle is not called, we complete $y$ with any $y_{| \{ i \} \times\{ j \}\times \{ k \} \times \mathbbm{Z}} \in \T_\Lang$.
\end{itemize}
This makes sure that $y$ is in the subshift $\Tfinal$, so $x\in\phi_{SA}(\Z e_1,\Tfinal)$.

\vspace{0.2cm}

The proof of Theorem is completed.\hfill{\tiny $\blacksquare$}

%\vspace{-0,2cm}

\paragraph{\textbf{An application of Theorem~\ref{EquivalenceOrdre}:}}
 There does not exist an ``universal'' subshift $\T$ which could simulate every element of $\shift$. Indeed, consider $\Lang=\Lang(\T)^c$, one has $\mathcal{C}l_{P,F,SA,SP,FT}(\T)=\{ \T_{\Lang'} : \Lang'\preceq \Lang\}$. But there exists $\Lang''$ strictly superior to $\Lang$ (see~\cite{rogersjr1987trf}). Moreover, one can choose $\Lang''$ such that for all patterns $p\in\Lang''\subseteq\E^d_{\A}$, then for all $p'\in\E^d_{\A}$ such that $p\sqsubset p'$, one has $p'\in\Lang''$. Thus $\Lang(\T_{\Lang''})^c=\Lang''$. One deduces that $\T_{\Lang''}\notin\mathcal{C}l_{P,F,SA,SP,FT}(\T)$.

\section*{Conclusion}

In this article we generalize the notion of tilings considering any set of forbidden patterns. We present operations on sets of tilings, called subshifts, inspired by the dynamical theory. We obtain different notions of simulation, depending on the set of operations which are considered. These notions involve different semi-orders on subshifts and in this article we focus on the semi-order which consider all the transformations presented. This semi-order is quite well understood since we establish a correspondence with a semi-order on languages of forbidden patterns based on computability properties. The following points are still open questions :
\begin{itemize}
 \item In our construction, considering two subshifts $\T_1$ and $\T_2$ respectively of dimension $d_1$ and $d_2$ such that $\Lang(\T_2)^c\preceq\Lang(\T_1)^c$, we need $\Tfinal\in\mathcal{C}l_{P,F,SA,SP,FT}(\T_1)$ of dimension $d_1+d_2+2$ to simulate $\T_2$. It is possible to decrease the dimension of $\Tfinal$?

 \item For which class $\Uclass\subseteq\shift$ there exists a subshift $\T$ such that $\mathcal{C}l_{P,F,SA,SP,FT}(\T)=\Uclass$?
\end{itemize}

We can also consider other semi-orders involved by other sets of operations and look for general tools to study them. In fact, some of these semi-orders have already been studied. For example, the set of space-time diagrams of a cellular automaton can be viewed as a subshift, and the orders presented in~\cite{mazoyer1999ioc,ollinger2003stacs,theyssier2005acm} could be formalized with the tools introduced in Section~\ref{operation}.

\bibliographystyle{alpha}
\bibliography{biblio}

\end{document}